\documentclass[prl,superscriptaddress,showpacs,twocolumn]{revtex4}
\usepackage{graphicx}
\usepackage{amssymb,amsmath}
\usepackage{subfigure}
\usepackage[dvips]{color}

\newcommand{\eref}[1]{Eq.~(\ref{#1})}
\newcommand{\fref}[1]{Fig.~\ref{#1}}

\newcommand{\bab}{%
           bonding/antibonding}

\newcommand{\bs}{%
           band-structure}

\newcommand{\im}{%
           \imath}
\newcommand{\bra}[1]{\ensuremath{\langle #1|}}
\newcommand{\ket}[1]{\ensuremath{|#1\rangle}}
\newcommand{\braket}[2]{\langle #1|#2\rangle}

\newcommand{\up}{%
        \ensuremath{\uparrow}}
\newcommand{\down}{%
        \ensuremath{\downarrow}}
        
\newcommand{\updown}{%
        \ensuremath{\uparrow\!\downarrow}}

\def\eg{\ensuremath{e_g^\sigma}}
\def\egp{\ensuremath{e_g^\pi}}

\def\t2g{\ensuremath{t_{2g}}}
\def\a1g{\ensuremath{a_{1g}}}

\def\se{self-energy}

\newcommand{\svek}{%
        \mathbf}
\newcommand{\vek}[1]{%
        \hbox{\textbf #1}}

\newcommand{\pr}{%
        ^\prime}

\def\vo2{VO\ensuremath{_2}}
\def\VO2{VO\ensuremath{_2}}
\def\v2o3{V$_2$O$_3$}
\def\V2O3{V$_2$O$_3$}
\def\vcro2{V$_{1-x}$Cr$_x$O$_2$}
\def\etal{{\it et~al.}}

\def\XXint#1#2#3{{\setbox0=\hbox{$#1{#2#3}{\int}$}
\vcenter{\hbox{$#2#3$}}\kern-.5\wd0}}

\begin{document}

\author{ Jan M. Tomczak}
\affiliation{Centre de Physique Th{\'e}orique, Ecole Polytechnique, CNRS, 
91128 Palaiseau Cedex, France}
\author{Ferdi Aryasetiawan}
\affiliation{Research Institute for Computational Sciences, AIST,
Umezono 1-1-1, Tsukuba Central 2, Tsukuba Ibaraki 305-8568, Japan}
\author{Silke Biermann}
\affiliation{Centre de Physique Th{\'e}orique, Ecole Polytechnique, CNRS,
91128 Palaiseau Cedex, France}

\title{
Effective \bs\ in the insulating phase versus\\ strong
dynamical correlations in metallic VO$_2$
}

\begin{abstract}
Using a general 
analytical continuation scheme for cluster
dynamical mean field calculations, we analyze real-frequency
self-energies, momentum-resolved spectral functions, and one-particle
excitations of 
the metallic and insulating phases of VO$_2$.
While for the former
dynamical correlations and
lifetime effects prevent a description in terms of 
quasi-particles,
the excitations 
 of the latter allow for an effective
\bs. 
We construct an orbital-dependent, but
static 
one-particle potential that reproduces the full many-body 
spectrum.
Yet, the 
 ground state is well beyond a static one-particle
description.
The emerging picture 
gives a non-trivial answer to the decade-old question of the nature of
the insulator, which we characterize
as a ``many-body Peierls'' state.
\end{abstract}

\maketitle

Describing electronic correlations is 
a challenge for
 modern condensed matter
physics.
While weak correlations slightly modify quasi-particle states,
by broadening them with lifetime effects
and shifting their 
energies, strong enough correlations
can entirely invalidate the 
 band picture by
inducing a Mott insulating state.

In a half-filled one-band model, an insulator is realized above a critical
ratio of 
interaction to 
 bandwidth. Though more complex scenarios exist in realistic
  multi-band cases,
a common feature of 
compounds
 that undergo a  metal-insulator
transition (MIT) 
upon the change of an 
external 
parameter, such as temperature or pressure, is that the respective
insulator 
 feels
  stronger correlations
than the metal, since it is precisely their
enhancement 
that drives the system
insulating.

In this paper we discuss 
 a material where
this rule of thumb is inverted~: We argue that in \vo2\
it is the insulator that is less correlated, in the sense that band-like
excitations are better defined and have longer lifetimes than in the
metal. Albeit, {\it neither} phase is
well
described by standard
 \bs\ techniques.
Using an analytical continuation scheme for 
quantum Monte Carlo solutions to Dynamical Mean Field Theory (DMFT)~\cite{tomczak_vo2_proc}, we discuss quasi-particle lifetimes, 
$\vek{k}$-resolved spectra (for comparison with future angle resolved
photoemission experiments) and effective \bs s.
While dynamical 
effects
 are crucial in the metal, the
excitations 
 of the insulator are
well described within a static picture~:
For the insulator 
we devise
an effective one-particle potential that
captures the interacting excitation spectrum. Still,
the corresponding 
ground state is far from a Slater
determinant, leading us
 to introduce the concept of a
``many-body Peierls'' insulator.

The MIT
 of \vo2\ has 
intrigued
solid state
physicists for
decades~\cite{PhysRevB.11.4383,goodenough_vo2,PhysRevLett.72.3389,sommers_vo2,pouget_review,PhysRevB.60.15699,eyert_vo2,korotin_vo2,tanaka_vo2,koethe:116402,eguchi_vo2,PhysRevB.41.4993,PhysRevB.20.1546}.
A high temperature metallic rutile (R)
phase transforms at T$_c$=340~K into an 
insulating monoclinic structure (M1), in which vanadium atoms pair up to
form tilted dimers along the c-axis.
The resistivity 
jumps up
 by two orders of magnitude, yet no local moments form. 
Despite extensive efforts, the mechanism of the transition
 is still
under 
debate~\cite{PhysRevLett.72.3389,eyert_vo2,eguchi_vo2, koethe:116402, PhysRevB.20.1546,tanaka_vo2,PhysRevB.41.4993}.
Two scenarios compete~: In the Peierls picture 
  the structural aspect (unit-cell doubling) 
   causes the MIT, while in the Mott picture
local correlations predominate. 

\vo2\ has a d$^1$ configuration and the crystal field splits the
3d-manifold into \t2g\ and empty \eg\ components. The
former further split into \egp\ and \a1g\ orbitals, which overlap in R-\vo2,
accounting for the metallic character.
Still, the quasi-particle peak seen in photoemission (PES)~\cite{koethe:116402,eguchi_vo2,PhysRevB.41.4993} is much
narrower than the Kohn-Sham spectrum of density functional theory (DFT) in the
local density approximation (LDA)~\cite{eyert_vo2}, and 
eminent
satellite features evidenced in PES are absent. 
In M1-\vo2, the \a1g\ form \bab\ orbitals, due to the 
dimerization.
As discussed by Goodenough~\cite{goodenough_vo2},
this also pushes up the \egp\ relative to the 
\a1g. 
Yet, the LDA~\cite{eyert_vo2}
yields
 a metal.
Non-local correlations beyond LDA were shown to be
essential~\cite{biermann:026404,liebsch:085109,laad:195120}.
Indeed, recent Cluster DMFT (CDMFT) calculations~\cite{biermann:026404}, in which
a two-site vanadium dimer constituted the DMFT impurity,
opened a gap, agreeing well with
 PES and x-ray experiments~\cite{PhysRevB.20.1546,koethe:116402}.

Starting from these LDA + CDMFT results~\cite{biermann:026404} for the 
Matsubara \t2g\ Green's
function $G(\im\omega_n)$
we 
deduce the
real frequency Green's function $G(\omega)$
 by the maximum entropy
method~\cite{maxent} and a
Kramers-Kronig transform.
The 
self-energy matrix $\Sigma(\omega)$
we obtain
by numerical inversion of
$G(\omega)=\sum_\svek{k}\left[\omega+\mu-H_\svek{k}-\Sigma(\omega)\right] ^{-1}$~\cite{tomczak_vo2_proc}, with the LDA Hamiltonian $H$, 
and the chemical potential $\mu$.
\begin{figure}
  \begin{center}

    \mbox{
      \hspace{-0.25cm}
      \subfigure{\scalebox{0.23}{\includegraphics[angle=-90,width=.9\textwidth]{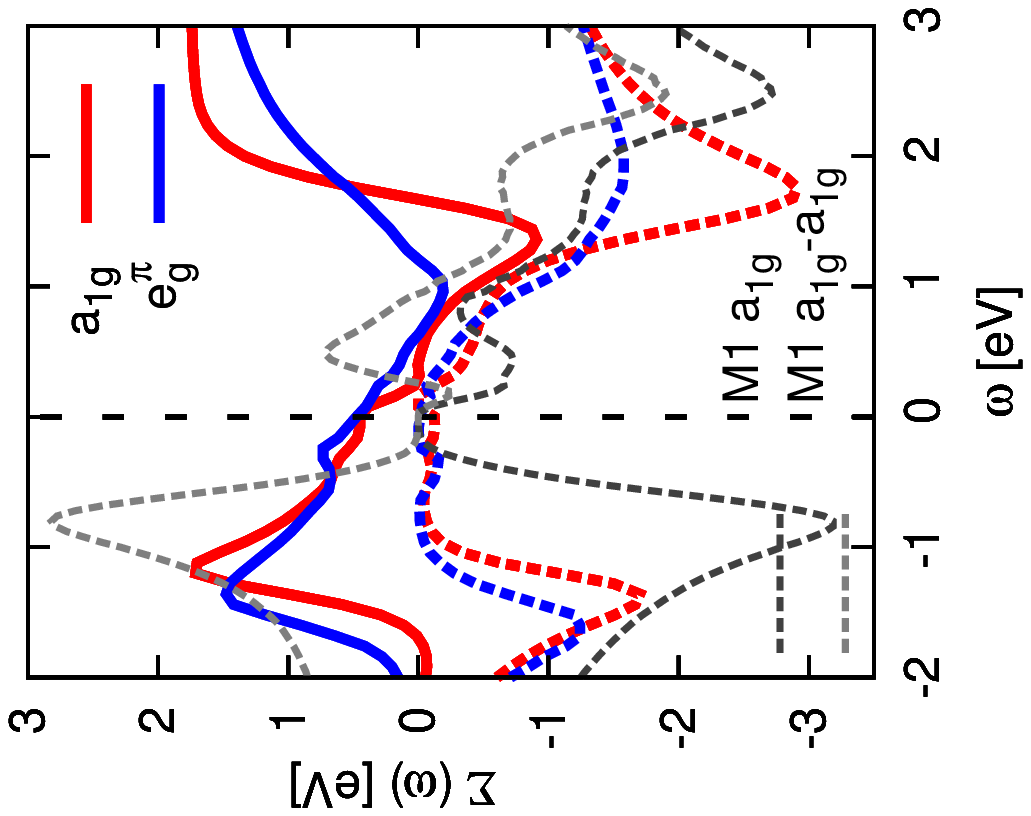}}} 
      \hspace{0.25cm}
      \subfigure{\hspace{-0.25cm}\scalebox{0.26}{\includegraphics[angle=-90,width=.9\textwidth]{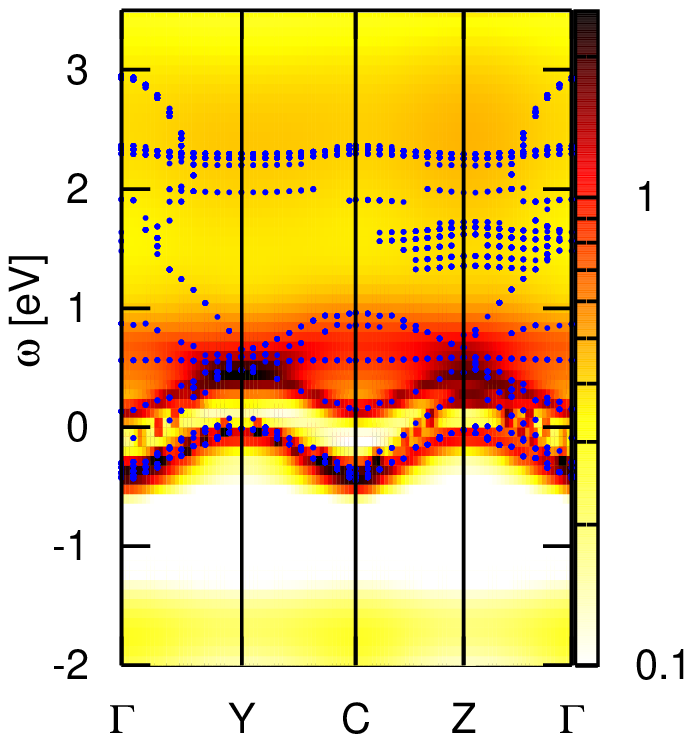}}} 
      }  
    \caption{(color online) Rutile VO$_2$~: (a) self-energy
      ($\Sigma-\mu$).
 Real (imaginary) parts are solid (dashed). As comparison
      M1 $\Im\Sigma_{\a1g}$, $\Im\Sigma_{\a1g-\a1g}$ are shown. (b)
      spectral function  A($\vek{k}$,$\omega$) and
      solutions of the QPE (blue). The LHB is the (yellow)
      region at -1.7~eV, the broad UHB appears (yellow) at $\sim$2.5~eV.}
    \label{fig1}
  \end{center}
\end{figure}

\fref{fig1} shows (a) the diagonal elements of the R-\vo2\ self-energy,
 and (b) the 
 resulting
  $\vek{k}$-resolved spectrum.
Notwithstanding minor
details, 
 the \a1g\ and \egp\
self-energies exhibit a 
similar {\it dynamical} behavior. 
The real-parts 
 at zero 
 energy, $\Re\Sigma(0)$, entailing relative
 shifts of
quasi-particle bands, are almost 
equal,
congruent with the low 
changes in their 
occupations vis-{\`a}-vis LDA~\cite{biermann:026404}, and with
the isotropy evidenced in experiment~\cite{haverkort:196404}.

Neglecting lifetime effects (i.e.\ $\Im\Sigma$$\approx$0), one-particle
excitations are given by  the poles of $G(\omega)$~:
$\det[\omega_\svek{k} + \mu - H_\svek{k} - \Re\Sigma(\omega_\svek{k})] =0$.
We shall refer to this as the quasi-particle equation (QPE)~\footnote{We solve the equation numerically by iterating
until self-consistency within an accuracy of 0.05~eV.}. For static or absent $\Re\Sigma$ this reduces to a simple eigenvalue problem.
In regions of 
 low $\Im\Sigma$, 
the QPE solutions
 will give an accurate description
of the position of spectral weight and constitute an effective \bs\ of the interacting system.
Yet, due to the frequency dependence, the number of solutions is no longer 
 bounded to the
number of orbitals.

Below (above) -0.5 (0.2)~eV,  the imaginary parts of the \se\ -- the inverse lifetime -- of R-\vo2\ is 
considerable.
Due to our limited precision for $\Im\Sigma(0)$, we have not attempted a
temperature dependent study to assess
 the experimental bad metal behavior, but the
resistivity
exceeding the Ioffe-Regel-Mott limit~\cite{qazilbash:205118} indicates
that even close to
the Fermi level, coherence is not fully reached.
At low energy, the QPE solutions (dots in \fref{fig1}b)
closely follow the spectral weight.
Above 0.2~eV, regions of high intensity appear,
howbeit, the larger $\Im\Sigma$
broadens the excitations, and no coherent features emerge, though the positions
of some 
\egp\ derived excitations
 are discernible. 
At high energies, positive and negative,
 distinctive 
features
 appear in $\Im\Sigma(\omega)$ 
 that are
responsible for 
lower (upper) Hubbard bands
(L/UHB), 
 seen in the 
spectrum
 at around -1.7 (2.5)~eV.
The UHB exhibits
a pole-structure
that reminds 
of the low-energy quasi-particle
\bs. 
Hence, an effective band picture is limited to the close
vicinity of the Fermi level, and
R-\vo2\ has to be considered as a strongly correlated metal (the
weight of the quasi-particle peak is of the order of 0.6).
This is experimentally
 corroborated 
  by the fact that an increase in the lattice spacing by
Nb-doping
results in a Mott insulator of 
rutile
 structure~\cite{pouget_review}.

\begin{figure}[t!h]
  \begin{center}
    \mbox{
\subfigure{\scalebox{0.23}{\includegraphics[angle=-90,width=.9\textwidth]{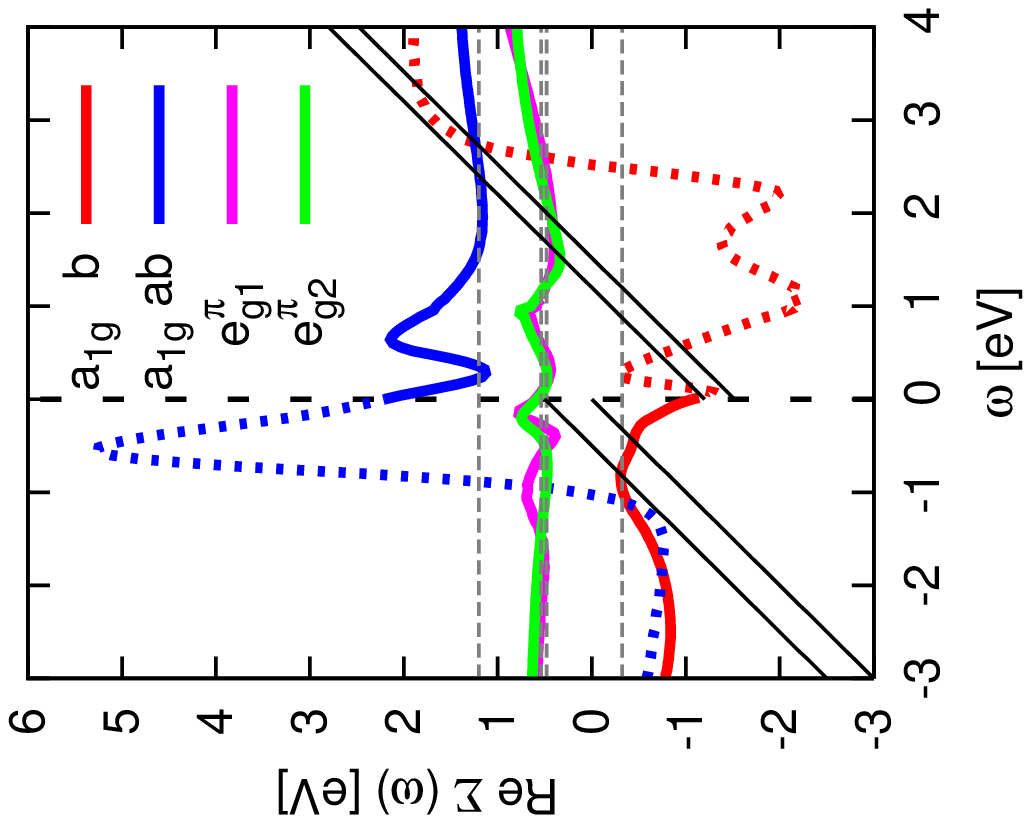}}} 
\hspace{.25cm}
\subfigure{\scalebox{0.23}{\includegraphics[angle=-90,width=.9\textwidth]{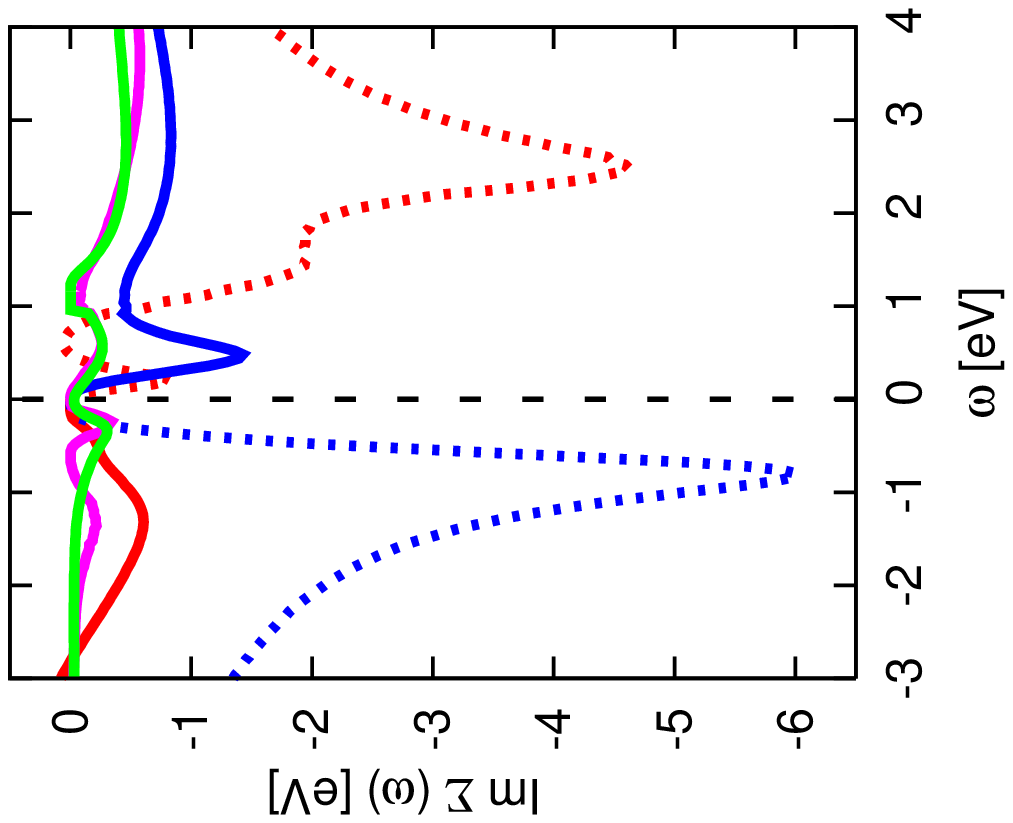}}} 
      }
    \caption{(color online) Self-energy ($\Sigma-\mu$) of M1-\vo2\ in the \a1g\ bab--basis~: (a) real parts. The black stripes delimit the \a1g\ LDA bandwidths, dashed horizontal lines indicate the values of the static potential $\Delta$.
 (b) imaginary parts. Self-energy elements are dotted in regions irrelevant for the spectrum.
    }
    \label{fig2}
  \end{center}
\end{figure}

\begin{figure}[!b!h!]
  \begin{center}
    \mbox{
      \hspace{-0.5cm}
      \subfigure{\includegraphics[clip=true,width=0.25\textwidth,angle=-90]{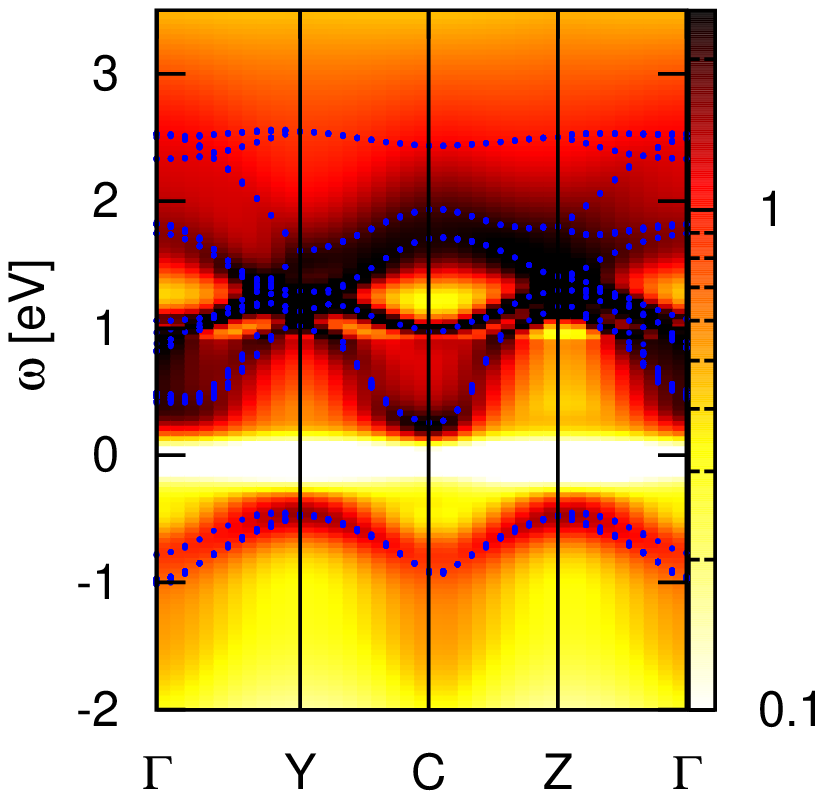}}
      \hspace{-0.25cm}
      \subfigure{\includegraphics[clip=true,width=0.25\textwidth,angle=-90]{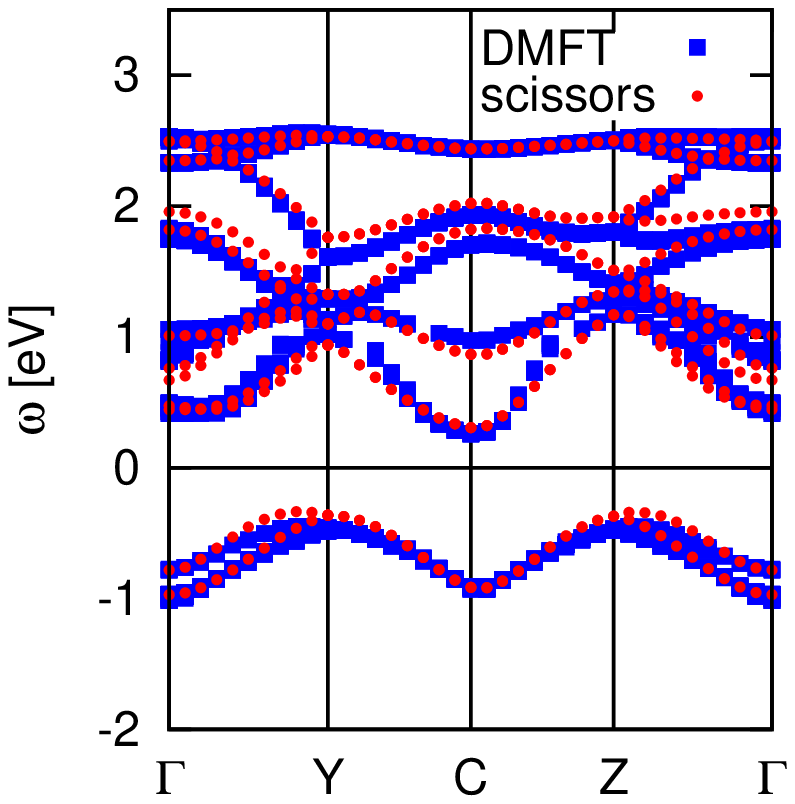}}
      }
    \caption{(color online) M1-\vo2~: (a) spectral
      function A($\vek{k}$,$\omega$).
    (blue) dots ((a) \& (b)) are solutions of the QPE. (b) The (red) dots
    are the eigenvalues of 
$H_\svek{k}$+$\Delta$. See text for discussion.
    }
    \label{fig3}
  \end{center}
\end{figure}

The imaginary parts of the M1 \a1g\
on-site, and \a1g--\a1g\ intra-dimer self-energies, \fref{fig1}a, are 
larger 
than 
in
R-\vo2, usually a
hallmark of 
increased 
correlations.
However, we shall argue 
that 
correlations
are in fact weaker than in the metal. 
Indeed, the dimerization in M1 leads to strong inter-site
fluctuations, 
evidenced
 by the significant intra-dimer \a1g--\a1g\ \se.
\fref{fig2} displays the M1-\vo2\ self-energy
in the \a1g\ \bab\ (bab) basis, 
$\Sigma_{b/ab}=\Sigma_{\a1g}\pm\Sigma_{\a1g-\a1g}$.
The \a1g\ (anti)bonding imaginary part 
 is low and varies little with frequency in the (un)occupied
part of the spectrum, thus allowing for coherent 
weight. 
In the opposite regions, the imaginary parts
reach huge values.
The \egp\ 
elements
are 
flat, and their imaginary parts tiny.
This is a direct consequence of the drastically reduced \egp\ occupancy
which drops to merely 0.14. 
These almost empty
orbitals 
feel only
weak correlations, and sharp bands are expected
at all energies.
A first idea for the \a1g\  
excitations is obtained
 from the intersections 
$\omega$+$\mu$$-\epsilon_{b/ab}(\svek{k})$=$\Re\Sigma_{b/ab}(\omega)$
as depicted 
in \fref{fig2}a, where the black stripes delimit
the LDA \a1g\ bandwidths.
The (anti)bonding band 
appears
 as the crossing
 of the (blue) red solid line with the stripe
at (positive) negative energy. Hence, the (anti)bonding band 
emerges
at (2.5) $-0.75$~eV.
Still, the antibonding band is much broadened since 
$\Im\Sigma_{ab}$ reaches
-1~eV.
To confirm 
this, 
we solved the QPE and calculated the $\vek{k}$-resolved spectrum (\fref{fig3}a). 
As expected,
 reasonably coherent weight appears over nearly the 
  {\it entire} 
   spectrum 
  from -1 to
  +2~eV, 
  whose position 
   coincides with the QPE poles~:
 The 
 filled 
bands 
correspond to
the \a1g\ bonding orbitals, while 
above the gap, 
 the \egp\ bands 
give rise to sharp 
features.
The anti-bonding \a1g\ 
 is not clearly distinguished
since \egp\ weight prevails
in this range.
The L/UHB have faded~:
a mere shoulder at -1.5~eV 
reminds of
the LHB. Finally,
contrary to R-\vo2, the number of poles equals the orbital
dimension.
Since, moreover,  the real-parts of the M1-\vo2\ \se\ 
 are almost constant
for relevant energies~\footnote{Explaining why LDA+U opens a
 gap~\cite{korotin_vo2,liebsch:085109},
yet while missing the correct 
\bab\ splitting.}, we construct a static potential,
$\Delta$, 
 by evaluating the dynamical \se\ 
at the LDA band centers (pole energies) for the \egp\ (\a1g), see
\fref{fig2}a~\footnote{
$\Delta_{\egp{}_{1}}$=$0.48$eV,
 $\Delta_{\egp{}_2}$=$0.54$eV, $\Delta_b$=$-0.32$eV,
 $\Delta_{ab}$=$1.2$eV}.
\fref{fig3}b shows the \bs\ 
 of $H_\svek{k}$+$\Delta$~:
 The agreement with the DMFT poles is excellent.
Our one-particle potential, albeit static,
depends on the orbital, and is thus non-local.
We emphasize the conceptual difference to the 
Kohn-Sham (KS) potential of DFT~:
The latter
generates
 an effective one-particle problem 
with the 
ground state density of
the true system. The KS
energies and states are {\it auxiliary} quantities.
Our one-particle potential, $\Delta$,
 on the contrary, was
designed to re\-produce the interacting excitations. 
The eigen\-values of $H_\svek{k}$+$\Delta$ 
are thus {\it not} artificial.
Still, like in DFT, the eigenstates are 
SDs by construction,
although the true states are not (see below).
The crucial point for M1-\vo2\ is that {\it spectral properties} are capturable
with this effective one-particle description.
 It is in this
sense that M1-\vo2\ exhibits
 only weak correlation effects.
The weight of the bonding excitation is
$Z$=$(1-\partial_\omega\Re\Sigma_b(\omega))^{-1}_{\omega=-0.7\hbox{\tiny
    eV}}$$\approx$0.75, and
thus larger than the rutile quasi-particle weight
(see above).

\begin{figure}[!b!h!]
  \begin{center}
    \vspace{-0.75cm}
    \mbox{
      \hspace{-0.325cm}
      \subfigure{\includegraphics[clip=true,width=0.315\textwidth,angle=-90]{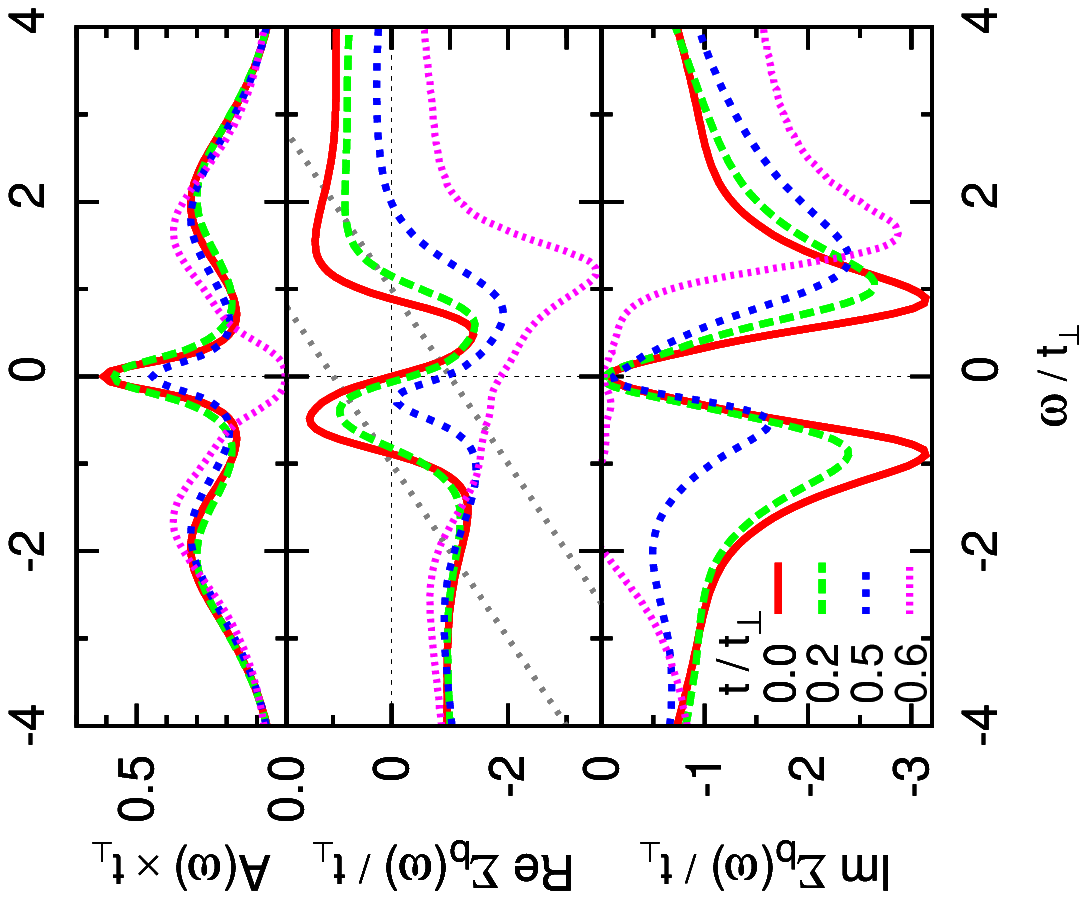}}
      \hspace{-0.5cm}
      \subfigure{\includegraphics[clip=true,width=0.315\textwidth,angle=-90]{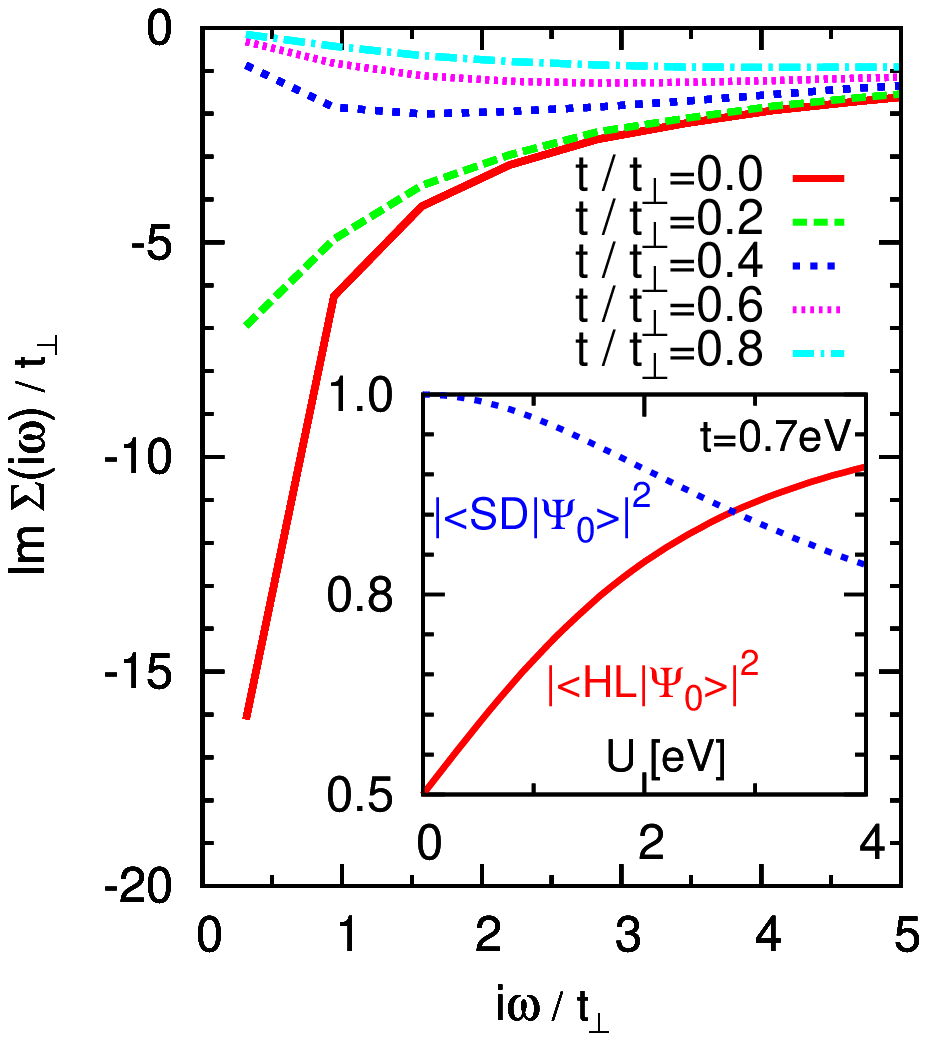}}
      }
    \caption{(color online)
		(a) spectral function (top), real (mid\-dle),
                imaginary (bottom) {\it bonding} \se\ $\Sigma_b$
                       of the CDMFT solution to \eref{model} for
                       $U$=4.0$t_\perp$, $\beta$=10/$t_\perp$, and
                       va\-rying intra-dimer hopping 
                       $t$/$t_\perp$.
                       $\Re\Sigma_b(\omega)$=$-\Re\Sigma_{ab}(-\omega)$,
                       $\Im\Sigma_b(\omega)$=$\Im\Sigma_{ab}(-\omega)$
                       by symmetry.
      (b) Imaginary Mat\-suba\-ra self-energy, $\Im\Sigma_b(\im\omega)$=$\Im\Sigma_{ab}(\im\omega)$, 
 for
$U$=6$t_\perp$, $\beta$=10/$t_\perp$ and varying $t$. 
      Inset: Projection of the SD and HL limit on the Hubbard
      molecule ground state ($t$=$0.7$~eV, $t_\perp$=0) versus U.
    }
    \label{fig4}
  \end{center}
\end{figure}

What is at the origin of this overall surprising coherence?
For the \egp\ orbitals, this simply owes to their depletion.
For the 
nearly
 half-filled \a1g\ orbitals the situation is more 
intricate.
It is a joint effect
of 
charge transfer into the
\a1g\ bands, and the \bab--splitting. Indeed, the filled bonding
band 
experiences
 only
weak fluctuations, 
due to its separation of several 
eV from the antibonding one.
To substantiate these qualitative arguments, we resort to 
the following model, 
which treats the solid as 
a collection
 of Hubbard dimers~: 
\begin{equation}\label{model}
H= -t\sum_{l\sigma}\left(
  c^\dag_{l1\sigma}c^{\phantom{\dag}}_{l2\sigma}\! +\! h.c.\right) -
  t_\perp \!\!\sum_{\stackrel{i=1,2}{\langle l,l\pr\rangle}}
c^\dag_{li\sigma}c^{\phantom{\dag}}_{l\pr i\sigma}
  +U\sum_{i l}n_{li\uparrow}n_{li\downarrow}
\end{equation}
Here, $c^\dag_{li \sigma}$ ($c_{li \sigma}$) creates 
(destroys)
 an electron with spin
$\sigma$ on site $i$ of the $l^{th}$ dimer.
 $t$ is the intra-dimer, $t_\perp$ the
inter-dimer hopping, $U$ the on-site Coulomb repulsion, and we assume half-filling. 
First, we 
discuss
 the $t_{\perp}$$\rightarrow 0$ limit, which 
is
an isolated dimer~: the Hub\-bard molecule.
We choose $t$=0.7~eV, the LDA intra-dimer 
\a1g--\a1g\ hopping, and $U$=4.0~eV~\cite{biermann:026404} for all
 evaluations.
The \bab--splitting,
$\Delta_{bab}$=$-2t
+ \sqrt{16t^2+U^2}$=3.48~eV, 
gets {\it enhanced} with respect to the $U$=0 case.
In M1-\vo2, the embedding
into the solid,
 and the hybridization with the \egp\
 reduce the splitting to $\sim$3~eV,
as can be inferred from the one-particle poles (\fref{fig3}), 
consistent
with experiment~\cite{koethe:116402}.
The ground state of the dimer is given by $\ket{\psi_0}=\left\{{4t}/\right.$$\left.(c-U)\right.$$\left.\left( \ket{\down\,\;\up}
-\ket{\up\,\;\down}\right)+\left(
 \ket{\updown\,\;0}
+ \ket{0\,\;\updown} \right)\right\}/a$~\footnote{$a=\sqrt{2\left({16t^2}/{(c-U)^2}+1\right)}$, $c=\sqrt{16t^2+U^2}$}
which is intermediate to the Slater determinant (SD) (the four states having equal weight), and the Heitler-London (HL) 
limit (double occupancies projected out). 
With the \vo2\ parameters,
the model dimer is close to the HL limit~\cite{sommers_vo2}.
The inset of \fref{fig4}b
shows the 
projections of the ground state onto the SD and the HL state. 
 The former, $\left|\braket{SD}{\psi_0}\right|^2$, 
equals the weight of the band-derived features in the spectrum (for $U$$>$0 satellites appear), while
the other measures the double occupancy $\sum_i\langle n_{i\uparrow} n_{i\downarrow} \rangle$ = 1 $-$ $\left|\braket{HL}{\psi_0}\right|^2$.
For $U$=4.0~eV the latter is 
largely
suppressed, 
as a consequence 
 of the interaction~:
The N-particle state is clearly not 
a SD. 
Still, the overlap with the SD, and thus the coherent weight, remains significant,
i.e.\ one-particle excitations survive and lifetimes are large.
To do justice to the seemingly opposing tendencies of correlation
driven non-SD-behavior, coexisting with a band-like 
 spectrum,
we introduce the notion of a ``many-body Peierls'' state. 

The charge transfer from the \egp\ into the then almost half-filled \a1g\ orbitals, 
finds its origin in the effective reduction of the local interaction in the bab--configuration~:
While for $U$=4~eV, 
    $\bra{SD} H \ket{SD}$ = 2.0~eV in the SD 
 limit, it
 reduces to merely 
$\bra{\psi_0} H \ket{\psi_0}$
= 0.91~eV in the ground state.
In fact, inter-site fluctuations are an
efficient way to avoid 
the on-site Coulomb repulsion. 
In M1-\vo2, this effect
manifests itself in a close cancellation of the
local and inter-site self-energies in the (un-)occupied
parts of the spectrum for the
(anti)bonding \a1g\ orbitals. 

The gap-opening in \vo2\ thus owes to two effects~:
The \se\ enhancement of
the \a1g\ bab--splitting, and
a charge transfer from the \egp\ orbitals.
The difference in $\Re\Sigma$ corresponds to this depopulation,
seen in experiments~\cite{haverkort:196404}
and theoretical studies~\cite{tanaka_vo2, biermann:026404}, and
leads to
the separation of the \a1g\ and \egp\
at the Fermi level. {\it The 
local interactions thus amplify Goodenough's scenario.}

To show that the embedding of the dimer into the solid does not
qualitatively alter our picture of the M1 phase, 
we solve the model, \eref{model}, using CDMFT.
This moreover allows to study the essentials of the rutile to M1 MIT by scanning through the degree of dimerization $t$ at constant interaction strength $U$ and embedding, or inter-dimer hopping, $t_\perp$. For the latter we assume 
a semi-circular density of states $D_\perp(\omega)$ of bandwidth $W$=4$t_\perp$.
In M1-\vo2, the  $t_\perp$ for direct \a1g-\a1g\ hopping is rather small, yet \egp-hybridizations lead to an effective $D_\perp$-bandwidth of about 1~eV.
We choose $U$=4$t_\perp$, and an inverse temperature $\beta$=10/$t_\perp$.
\fref{fig4}a
displays 
 the orbital traced local spectral function
 $A(\omega)$=$A_b(\omega)+A_{ab}(\omega)$ 
 (b,ab denoting again the \bab\ combinations)
  and the bonding \se\ $\Sigma_b(\omega)$ for different 
  intra-dimer hoppings $t$~: In the absence of $t$, the result equals by construction the
single site DMFT solution ($\Sigma_b$=$\Sigma_{ab}$), which, for our
parameters, is a correlated metal, 
analog to R-\vo2.
 The spectral weight at the Fermi level is given by $A_{b/ab}(0)=D_\perp(\pm t-\Re\Sigma_{b/ab}(0))$, with $\Re\Sigma_{b/ab}(0)$=$\mp\,\Re\Sigma_{ab}(0)$.
Thus a MIT occurs at $t+\Re\Sigma_{ab}(0)=2t_\perp$, 
 when all spectral weight has been shifted out of the bandwidth~: Above $t$/$t_\perp$$=$$0.5$ we find a many-body Peierls phase corresponding to M1-\vo2.
In \fref{fig4}a we have indicated again the graphical QPE
approach~: The system evolves from three solutions {\it per orbital}
(Kondo resonance, L/UHB) at $t$=0 to a single one at $t$/$t_\perp$=0.6. Hence
the peaks in the insulator are {\it not} Hubbard satellites, but just shifted bands. 
The embedding, $t_\perp$, broadens the excitations and washes out
the satellites of the isolated dimer, 
like for M1-\vo2. 
Still, as a function of t, the coherence of the spectrum increases,
since the imaginary part of the (anti-)bonding \se\ subsides at
the renormalized (anti-)bonding excitation energies.
Our model thus captures the essence of the rutile to M1
transition, reproducing both, the dimerization induced increase in
coherence, and the shifting of excitations.

Under uni-axial pressure or Cr-doping, \vo2\ develops the insulating M2 phase~\cite{pouget_review} in which
every se\-cond vanadium chain along the c-axis consists of untilted dimers, whereas in the others only the tilting occurs.
We may now speculate that the 
dimerized pairs in M2 form \a1g\ Peierls singlets as in M1, while the
tilted pairs are in a Mott state.
Hence,
 we interpret the seminal work
of~\cite{pouget_review} 
 as the observation of a
 Mott to many-body Peierls transition  
taking place on the tilted chains when
going from M2 to M1.
To illustrate this, 
we solve again \eref{model}  for appropriate parameters.
The tilted M2 chains are akin to the rutile phase, yet with a reduced
\a1g\ bandwidth~\cite{eyert_vo2}. Thus we now choose
$U$=6$t_\perp$, $\beta$=$10$/$t_\perp$, and vary $t$.
All solutions shown in \fref{fig4}b are insulating, however, the diverging self-energy  at vanishing intra-dimer coupling
 ($t$=0, tilted ``M2'' chains) 
becomes regularized with the bond enhancement ($t$$>$0, ``M1''). The 
imaginary part of the \se\ gets flatter and the system thus more coherent.
The above is consistent with the finding of (S=0) S=1/2 for the (dimerized) tilted
pairs in M2-\vo2~\cite{pouget_review}.

While our results do not exclude surprises in the direct vicinity of
T$_c$~\cite{kim:266401},
the nature of insulating \vo2\ is shown to be rather
``band-like'' in the above sense.
Our analytical continuation scheme allowed us to {\it explicitly
calculate} this \bs.
The latter 
can also be derived from a static
one-particle potential.
Yet, this does {\it not} imply a one-particle picture for quantities
other than the spectrum.
Above all,
the ground state is not a Slater determinant.
Hence, we qualify M1-\vo2\ as a ``many-body Peierls'' phase.
We argue that the 
 weakness
of lifetime effects results from strong inter-site fluctuations
that circumvent 
local interactions
 in an otherwise strongly correlated solid.
This is in striking contrast to the strong dynamical correlations in
the metal, which is dominated by important lifetime effects and 
 incoherent features.

We thank H. T. Kim, J. P. Pouget, M. M. Qazilbash, and A. Tanaka for 
valuable
discussions and A. I. Poteryaev, A. Georges and A. I. Lichtenstein for
discussions
 and
the collaboration~\cite{biermann:026404} that was 
our starting point.
We 
thank AIST, Tsukuba, for hospitality. 
JMT was supported by a JSPS fellowship.
Computer time was provided by IDRIS, Orsay (project No. 071393).


\bibliographystyle{apsrev}

\end{document}